\documentclass[secnumarabic,aps,prb,reprint,superscriptaddress,floatfix]{revtex4-2}
\usepackage[english]{babel}
\usepackage{graphicx}
\usepackage{color}
\usepackage{amsmath}
\usepackage{amssymb}
\usepackage{tabularx}
\usepackage[dvipsnames]{xcolor}
\usepackage{natbib}

\begin{document}

\title{Orbital correlations in ultrathin films of late transition metals}
\author{Sergei Ivanov}
\author{Joshua Peacock}
\author{Sergei Urazhdin}
\affiliation{Department of Physics, Emory University, Atlanta, GA, USA.}

\begin{abstract}

We develop a two-orbital Hubbard model of electron correlations in ultrathin (111)-oriented fcc films of late transition metals such as Co and Ni. Our model indicates that the Mott-Hund's interaction results in ferromagnetic nearest-neighbor orbital correlations. Frustration associated with the mismatch between orbital and crystal symmetries prevents orbital ordering, resulting in the orbital liquid state. This state can be manifested in phenomena involving spin-orbit coupling, such as magnetic anisotropy.
\end{abstract}
\maketitle

\section{Introduction}

Magnetism and superconductivity are among the most fascinating collective phenomena in condensed matter physics. These phenomena stem from electron correlations that provide a mechanism for electrons to lower their kinetic and Coulomb interaction energies while satisfying the Pauli exclusion principle~\cite{kittel1987quantum}. Despite this fundamental connection, the two phenomena are usually approached from very different perspectives. Superconductivity is understood in terms of pairwise electron correlations (Cooper pairs) that cannot be described in single-particle terms~\cite{tinkham2004introduction,Tilley2019}. 

In contrast, magnetism is commonly analyzed in single-particle terms. For instance, in the Stoner-Weiss model, ferromagnetic (FM) ordering arises because electrons with the same spin avoid each other due to the Pauli principle, reducing their Coulomb interaction energy~\cite{Stoner1937,Stoner2008}. Fundamentally, this is a many-particle effect. Nevertheless, it can be captured in the single-particle mean-field approximation as an effective exchange field~\cite{handley2000modern}. It is notable that the local density approximation (LDA) of the {\it ab-initio} theory, which includes this effect as the single-particle exchange-correlation energy, generally needs to be amended with additional correlation terms to adequately describe FM systems~\cite{https://doi.org/10.1002/qua.24521}, indicating that correlations not reducible to single-particle energies may play an important role in magnetism.

\begin{figure}
	\includegraphics[width=0.5\linewidth]{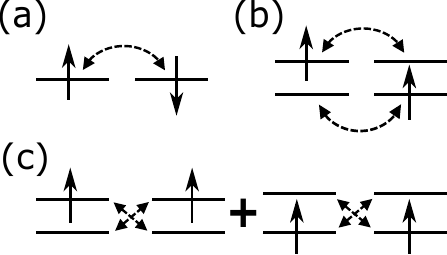}
	\caption{(a) Mott mechanism of AFM spin ordering in single-orbital systems. (b) Kugel-Khomskii mechanism of FM spin ordering in two-orbital systems dominated by same-orbital hopping. (c) Orbital correlations in FM spin-ordered state of ultrathin late transition metal films identified in our model. The double-headed dashed arrows show virtual hopping.}\label{fig:MKK}
\end{figure}

A prominent example of a model of magnetism based on such correlations is the Mott theory of antiferromagnetic (AFM) ordering in a single half-filled band~\cite{Fazekas1999-cr}. The kinetic energy of two electrons on neighboring sites is minimized if their spins are AFM-correlated so that the Pauli principle allows hopping onto the neighboring site, stabilizing an AFM-ordered insulating ground state (g.s.) at sufficiently large interaction, Fig.~\ref{fig:MKK}(a). 

Kugel-Khomskii model is a conceptually similar theory of ferromagnetism in cubic magnetic insulators characterized by two degenerate orbitals populated with one electron per site~\cite{Kugel1982}. Same-orbital hopping of an electron to the neighboring site occupied by another electron results in the lowest Mott-Hund's onsite energy if electrons are on different orbitals and have the same spin, stabilizing FM spin ordering and AFM orbital ordering [Fig.~\ref{fig:MKK}(b)]. Despite its success in describing cubic FM insulators, a similar model for common transition metal ferromagnets has not yet emerged.

In this work, we utilize the Hubbard model to analyze correlations in (111)-oriented ultrathin fcc films of late transition metals such as Co and Ni. Our choice of this system is motivated by the relatively simple electronic structure characterized by the d-level population of about one hole per site, enhanced correlation effects due to hopping suppression in thin films, and practical relevance for magnetic memory and sensor applications~\cite{doi:10.1063/1.2838228}.

Our main finding is illustrated in Fig.~\ref{fig:MKK}(c). Opposite-orbital hopping is dominant, resulting in FM orbital coupling between neighboring sites. Virtual hopping stabilizes an orbital singlet state - a superposition of FM-coupled orbital states of neighboring sites characterized by vanishing orbital angular momentum on each site. In contrast to the Kugel-Khomskii model, orbital ordering is prevented by orbital frustration due to the mismatch between orbital and crystal symmetries. The resulting state can be described as an orbital liquid~\cite{Khaliullin2000-xe}. 

We also show that spin-orbit coupling (SOC) breaks the symmetry between the orbital singlet components, resulting in a finite orbital moment that facilitates perpendicular magnetic anisotropy (PMA). This finding sheds new light on the enhanced PMA commonly observed in ultrathin magnetic films~\cite{Bruno1989-zn,Yu2018-ef} and suggests an approach to efficient control of PMA in magnetic nanodevices via electron correlations.

The paper is organized as follows. In the next section, we introduce our model. In Section~\ref{sec:2s}, we show that the two-site approximation is equivalent to the Kugel-Khomskii model, aside from an inconsequential hopping asymmetry. In Section~\ref{sec:3s}, we utilize a three-site model to elucidate orbital frustration associated with the mismatch between orbital and lattice symmetries, which prevents orbital ordering in this system. In Section~\ref{sec:extended}, we extrapolate our analysis to an extended system. In Section~\ref{sec:SOC}, we show that orbital correlations can be manifested in phenomena involving SOC, focusing on magnetic anisotropy as a specific example. Finally, in Section~\ref{sec:discussion} we summarize our results and discuss their broader implications for other magnetic systems expected to exhibit similar orbital frustration.

\begin{figure}
	\includegraphics[width=0.6\linewidth]{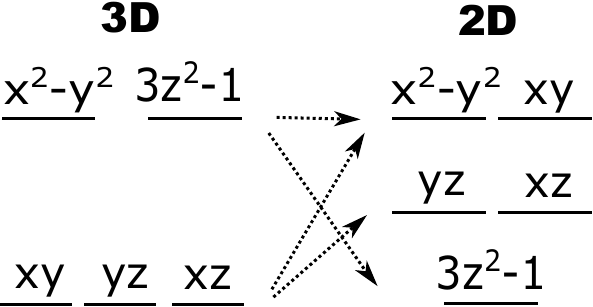}
	\caption{Splitting of atomic d-levels by the cubic crystal-field symmetry of a three-dimensional fcc crystal (left) and the $C_6$ symmetry of the two-dimensional triangular lattice (right). The vertical scale does not represent level energies, which are separately discussed in the text.}\label{fig:d-levels}
\end{figure}

\section{Model}\label{sec:model}

The magnetic and electronic properties of transition metals are dominated by the states derived from the 4s and 3d atomic orbitals. Strong hybridization between 4s orbitals produces a broad ($\approx10$~eV) 4s band~\cite{handley2000modern}. In contrast, the 3d band is significantly narrower and is formed by five sub-bands derived from the corresponding atomic 3d orbitals. Ferromagnetism mediated in transition metal ferromagnets by the hybridization of d-orbitals is a consequence of direct exchange interaction between d-electrons, as opposed e.g. to indirect exchange between localized f-electrons in rare-earth ferromagnets mediated by the conduction electrons~\cite{White2007}. This interaction is commonly approximated by the Heisenberg exchange Hamiltonian $\hat{H}=-J\sum_{\left<i,j\right>}\hat{S}_i\hat{S}_j$ describing coupling between spins $\hat{S}_i$, $\hat{S}_j$ of d-electrons quasi-localized on the neighboring lattice sites. Since 4s states do not play a significant role in magnetic ordering, we neglect them in our analysis focused on the mechanisms of magnetism.

To identify the orbitals that dominate valence states in ultrathin (111)-oriented fcc films, we consider the d-level splitting by the effective crystal field~\cite{PhysRev.41.208}. In bulk transition metals with cubic symmetry, the crystal field splits the five-fold degenerate atomic d-level into three-fold orbitally degenerate $t_{2g}$ orbitals and two-fold degenerate $e_{g}$ orbitals [Fig.~\ref{fig:d-levels}, left]~\cite{handley2000modern}. In (111)-oriented ultrathin fcc films of Ni or Co with a thickness comparable to the lattice constant, bonding is dominated by the six nearest in-plane neighbors on the triangular lattice. In the two-dimensional approximation, the six-fold rotational symmetry of the crystal field splits the d-levels of an atom into an orbitally non-degenerate level, $d_{3z^2-1}$, and two doubly orbitally degenerate levels formed by the orbitals $d_{yz}$, $d_{xz}$ and $d_{x^2-y^2}$, $d_{xy}$, respectively [Fig.~\ref{fig:d-levels}, right]~\cite{Willock}. 

The hole representation provides a simple way to identify the orbitals that dominate valence states. In Ni, about half an electron per site resides in the 4s shell, leaving about half a hole in the d-band~\cite{handley2000modern}. In Co, there are about $1.5$ holes per site in the d-band. Thus, only the bonding states derived from the most strongly hybridized 3d orbitals are occupied with holes. The strongest hybridization is expected for the $d_{x^2-y^2}$ and $d_{xy}$ orbitals characterized by the largest in-plane extent. We have confirmed this qualitative conclusion by analyzing the hybridization using the Koster-Slater parameters corresponding to the interatomic spacing of Co and Ni~\cite{Harrison}.

We describe the system by the Hubbard-Kanamori Hamiltonian in the hole representation,
\begin{widetext}
\begin{equation}\label{eq:Hunds}
    \begin{split}
    \hat{H}&=\sum_{\vec{n},\vec{l},\sigma,\sigma',s} t_{\vec{n},\vec{n}+\vec{l}}^{\sigma,\sigma'}\hat{c}^{\dagger}_{\vec{n}+\vec{l},\sigma',s}\hat{c}_{\vec{n},\sigma,s} +
    U\sum_{\vec{n},\sigma}\hat{n}_{\vec{n},\sigma,\uparrow}\hat{n}_{\vec{n},\sigma,\downarrow}+U'\sum_{\vec{n},\sigma}\hat{n}_{\vec{n},\sigma,\uparrow}\hat{n}_{\vec{n},-\sigma,\downarrow}
    +U''\sum_{\vec{n},s}\hat{n}_{\vec{n},+,s}\hat{n}_{\vec{n},-,s}+\\
    &+J\sum_{\vec{n},\sigma}\hat{c}^{\dagger}_{\vec{n},\sigma,\uparrow}\hat{c}^{\dagger}_{\vec{n},-\sigma,\downarrow}\hat{c}_{\vec{n},\sigma,\downarrow}\hat{c}_{\vec{n},-\sigma,\uparrow}
		+J_c\sum_{\vec{n},\sigma}\hat{c}^{\dagger}_{\vec{n},\sigma,\uparrow}\hat{c}^{\dagger}_{\vec{n},\sigma,\downarrow}\hat{c}_{\vec{n},-\sigma,\downarrow}\hat{c}_{\vec{n},-\sigma,\uparrow} +\sum_{\vec{n},s,s',\sigma,\sigma'}\lambda_{s,s'}^{\sigma,\sigma'} \hat{c}^{\dagger}_{\vec{n},\sigma,s}\hat{c}_{\vec{n},\sigma',s'},
	\end{split}
\end{equation}
\end{widetext}
where the summation is over all the dummy indices, $\hat{c}^{\dagger}_{\vec{n},\sigma,s}$ creates a hole with spin $s=\pm1/2\equiv\uparrow$, $\downarrow$ and pseudospin $\sigma=\pm 1$ enumerating the orbitals $d_{x^2-y^2}$, $d_{xy}$ at the position $\vec{n}$ normalized by the lattice constant, $\vec{l}$ is a unit vector in the direction of one of the six nearest neighbors, and $\hat{n}_{\vec{n},\sigma,s}=\hat{c}^{\dagger}_{\vec{n},\sigma,s}\hat{c}_{\vec{n},\sigma,s}$. Note that $\vec{n}$ is generally not an integer vector for the triangular lattice.

The first term in Eq.~(\ref{eq:Hunds}) accounts for the intra- and inter-orbital nearest-neighbor hopping described by the matrix elements $t_{\vec{n},\vec{n}+\vec{l}}^{\sigma,\sigma'}$. The next three terms describe onsite Coulomb interactions between holes with opposite spins in the same orbital, opposite spins in different orbitals, and the same spin in different orbitals, respectively, with the coefficients $U\ge U'>U'' >0$ accounting for the atomic Hund's rules.

The two terms with coefficients $J$ and $J_c$, commonly referred to as the Kanamori's spin-flip and pair-hopping terms, are required by the symmetry of the Hamiltonian. The last term accounts for SOC, which controls the magnetic anisotropy essential for the magnetic devices based on thin films with PMA~\cite{Poletkin2022-hd}. Since the magnitude of SOC is at least an order of magnitude smaller than the dominant hopping and interaction energies, it is neglected in the next two sections focused on magnetic ordering and is separately discussed in Section~\ref{sec:SOC}.

The symmetry-constrained relations among the parameters $U, U', U'', J$, and $J_c$ in Eq.~(\ref{eq:Hunds}) are $U=U_0+J_0/2$, $U'=U_0-J_0/2$, $U''=U_0-J_0$, $J=J_c=J_0/2$ in the orbital basis of cubic harmonics, and 
$U=U'=U_0$, $U''=U_0-J_0$, $J=J_0$, $J_c=0$ in the basis of spherical harmonics $d_{\pm2}=(d_{x^2-y^2}\pm id_{xy})/\sqrt{2}$. We use the values $U_0=3.64$~eV, $J_0=0.77$~eV based on extensive prior Hubbard modeling of transition metal compounds~\cite{pavarini2017the,PhysRevB.83.205112}. We do not include Zeeman energy produced by the magnetic field, whose main role away from the critical points is to control the direction of magnetization.

\section{Two-site model}\label{sec:2s}

In this section, we consider a two-site approximation, and show that in the cubic harmonic basis the nearest-neighbor orbital correlations are essentially the same as in the Kugel-Khomskii model. In the subsequent sections, we show that the form of correlations becomes different in the spherical harmonic basis, which is more practical for analyzing extended systems and SOC. We choose the x-axis along the direction between the two sites, the y-axis in-plane perpendicular to it, and the z-axis normal to the plane. Intra-orbital hopping between sites is described by the matrix elements $t_{x^2-y^2}$ and $t_{xy}$. Inter-orbital hopping vanishes because the orbital $d_{x^2-y^2}$ is symmetric with respect to the inversion of the y-axis, while the orbital $d_{xy}$ is antisymmetric. Using the Koster-Slater parameters, and accounting for the fact that the signs of hopping amplitudes in the hole representation are opposite to those in the electron representation, we calculate $t_{x^2-y^2}=0.32$~eV, $t_{xy}$ $=-0.27$~eV for Ni, and $t_{x^2-y^2}=0.39$~eV, $t_{xy}$ $=-0.33$~eV for Co with a precision of about $0.01$~eV~\cite{Harrison}. The only distinction between this system and the system considered in the Kugel-Khomskii model is a small difference between the amplitudes of hopping on the two orbitals.

\begin{figure}
	\includegraphics[width=0.7\linewidth]{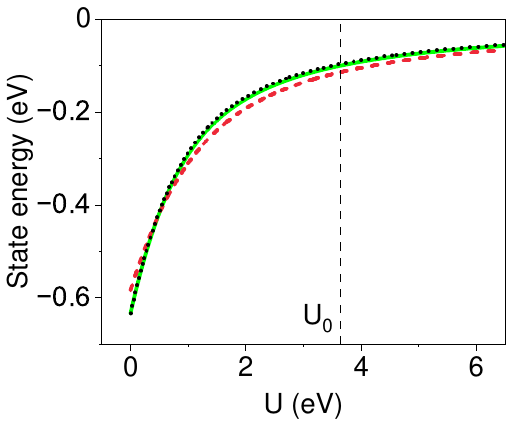}
	\caption{State energies vs $U$ for the two-site model, with $J_0=0.21U_0$. FM (spin triplet, red dashed line) and nonmagnetic (spin singlet, black dots) state energies calculated analytically as described in the text, solid green curve: result of exact numeric diagonalization for the singlet. Dashed vertical line marks the value of the Mott energy used in our model. The calculations are based on the hopping parameters of Ni.}\label{fig:two_sites}
\end{figure}

To analyze correlations, we consider the system populated with two holes. In the limit of negligible interaction, the two-particle g.s. is the single-particle g.s. populated with two opposite-spin holes. At finite interaction, this state evolves into a singlet whose energy calculated approximately by neglecting the last two interaction terms in Eq.~(\ref{eq:Hunds}) is $E^{(2)}_0=U/2-\sqrt{(U/2)^2+4t^2_{x^2-y^2}}$. The effects of these interaction terms calculated perturbatively result in energy correction $\Delta E=J_0E^{(2)}_0/(4E^{(2)}_0-2U_0)$, in good agreement with exact numeric diagonalization, Fig.~\ref{fig:two_sites}.

We now show that at sufficiently large interaction the g.s. is a spin triplet. The qualitative picture is the same as in the Kugel-Khomskii model. In the regime dominated by interactions, one can treat hopping as a perturbation on the manifold of degenerate ground states of interaction Hamiltonian with one hole per site. The second-order correction to the energy of this state due to hopping is minimized if the Mott-Hund's interaction energy of two onsite holes produced by virtual hopping is minimized, i.e. they have the same spin and are on the opposite orbitals. Thus, the spins of the holes on the two sites are FM coupled and their orbitals are AFM coupled, consistent with Fig.~\ref{fig:MKK}(b). 

This picture is confirmed by the exact diagonalization of the Hamiltonian. For simplicity, we consider only two components of spin triplet with spin normal to the film, whose wavefunctions labeled by the spin direction $s$ are 
\begin{equation}\label{eq:spin_2s}
	\begin{split}
		\psi^{(2)}_{t,s}&=[\frac{\cos\theta_t}{\sqrt{2}}(\hat{c}^{\dagger}_{1,x^2-y^2,s}\hat{c}^{\dagger}_{2,xy,s}-\hat{c}^{\dagger}_{2,x^2-y^2,s}\hat{c}^{\dagger}_{1,xy,s})\\
		&-\frac{\sin\theta_t}{\sqrt{2}}(\hat{c}^{\dagger}_{1,x^2-y^2,s}\hat{c}^{\dagger}_{1,xy,s}-\hat{c}^{\dagger}_{2,x^2-y^2,s}\hat{c}^{\dagger}_{2,xy,s})]|0\rangle
	\end{split}
\end{equation}
with energy $E_t=U''/2-\sqrt{(U''/2)^2+(t_{x^2-y^2}-t_{xy})^2}$. Here, we use trigonometric parameterization $\theta_t=\tan^{-1}[-E_t/(t_{x^2-y^2}-t_{xy})]$ of the amplitudes to simplify normalization. For the hopping parameters of Ni, $E_t=-0.2$~eV, $\theta_t=14^\circ$ and this state is dominated by the first term. Thus, the holes are almost localized, consistent with the qualitative analysis above.

{\it Stoner vs Heisenberg magnetism.} We now discuss the contribution of interaction-induced correlations between holes to FM ordering in the considered two-site model. At sufficiently small interaction, the FM (spin triplet) state is approximated as a product of single-particle states with the same spins, which can be viewed as the two-site limit of Stoner (single-particle band) magnetism. In this regime, the energies of both the singlet and the triplet linearly vary with $U$ [see Fig.~\ref{fig:two_sites}, $U\lesssim 1$~eV]. The smaller slope for the FM state is a consequence of the Pauli principle, which allows same-spin particles to avoid each other and lower their interaction energy relative to the spin singlet state. However, the dependence significantly deviates from linear at the experimentally relevant values of $U$, suggesting that the Stoner approximation is not adequate for this model system.

Instead, in the experimentally relevant regime at $U=U_0$, the first two components in Eq.~(\ref{eq:spin_2s}), which describe two holes localized on different sites, dominate the FM wavefunction, which places this system in the strongly correlated limit~\cite{Fazekas1999-cr,coleman_2015}. This regime can be described as the Hund's-Heisenberg magnetism, in which
 FM coupling between spins of quasi-localized particles is associated with the onsite Hund's interaction mediated by virtual hopping. We note that the spin-dependence of Hund's interaction itself is associated with the same Pauli exclusion principle that underlies Stoner magnetism. The difference is that in the Hund's-Heisenberg mechanism, the spin-dependence of many-electron energy is governed by the interaction-induced electron correlations neglected in the Stoner mechanism.

\begin{figure}
	\includegraphics[width=0.9\linewidth]{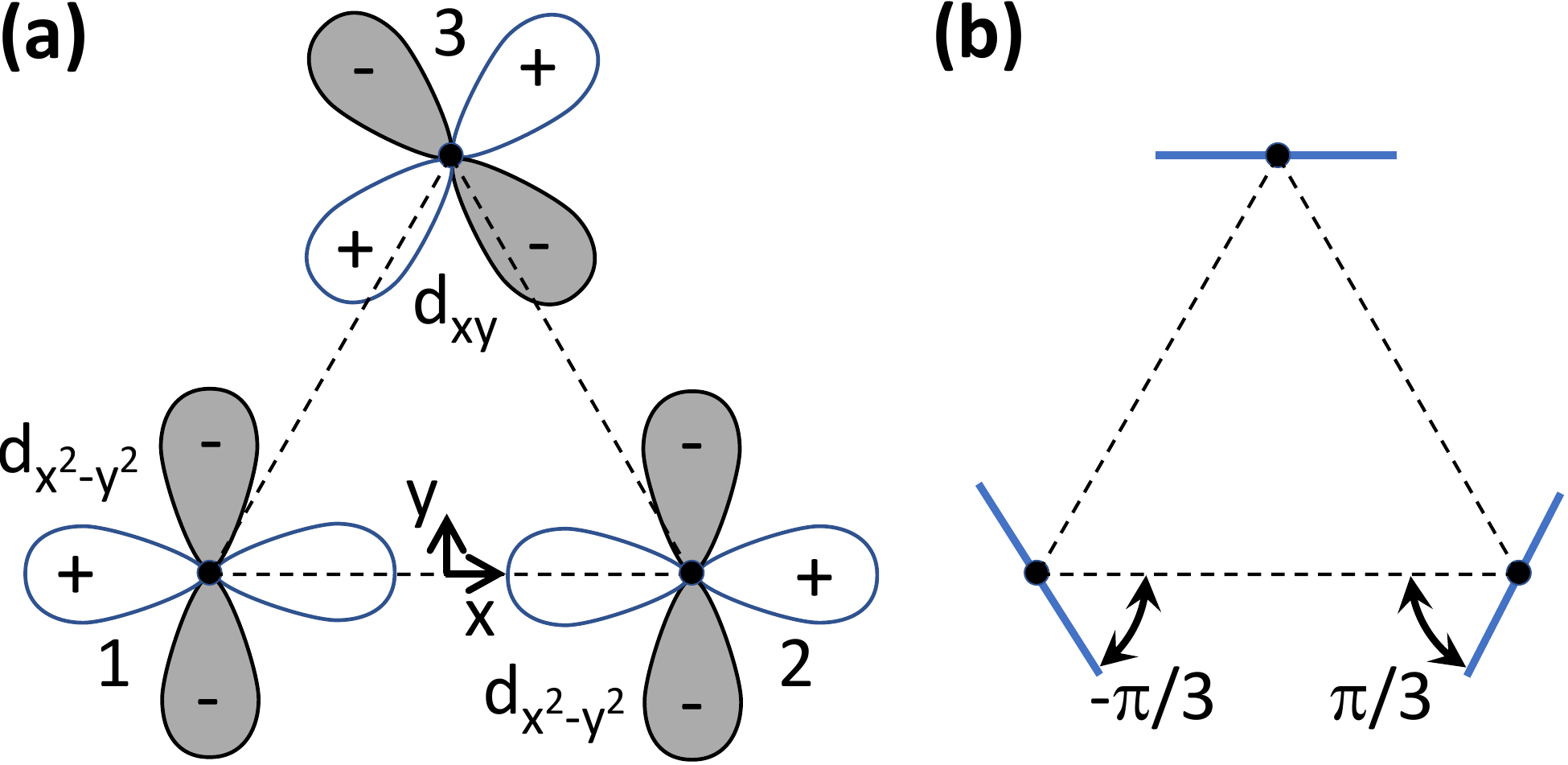}
	\caption{Three-site model. (a) Cubic harmonic basis. Orbital $d_{x^2-y^2}$ is shown on sites $1$ and $2$, and orbital $d_{xy}$ is shown on site $3$. (b) Rotated spherical harmonic basis. Solid lines show the directions along which the basis harmonics defined in the text are real and positive.}\label{fig:three_sites}
\end{figure}

\section{Three-site model of FM state}\label{sec:3s}

In this section, we show that the similarity to the Kugel-Khomsii model does not extend beyond the two-site approximation, due to the orbital frustration associated with the mismatch between orbital and lattice symmetries. For the square lattice, the matrix elements $t_{x^2-y^2}$, $t_{xy}$ would describe orbitally-selective hopping between a given site and all four of its nearest neighbors, due to the matching symmetries of the lattice and the $d$-orbitals. This would allow for a straightforward extension of the two-site model to an extended system, leading to orbital ordering as in the Kugel-Khomskii model~\cite{Kugel1982}. However, the fcc lattice of bulk late transition metals or the triangular lattice of (111)-oriented ultrathin films do not permit such an extension. The three-site model discussed in this section illustrates the resulting orbital frustration effects, providing a step toward the model of the extended system discussed in the next section.

We position sites $1$ and $2$ on the x-axis, the same way as in the two-site model, and site $3$ on the y-axis [Fig.~\ref{fig:three_sites}(a)]. In this configuration, the orbitally selective matrix elements $t_{x^2-y^2}$, $t_{xy}$ introduced in the two-site model describe hopping between sites $1$ and $2$, but not between one of these two sites and site $3$. For instance, the lobes of the $d_{xy}$ orbital on site $3$ almost face the lobes of the orbitals $d_{x^2-y^2}$ on sites $1$ and $3$, resulting in finite inter-orbital hopping characterized by direction- and site-dependent matrix elements. This difficulty has been discussed in the context of the compass models of anisotropic spin coupling~\cite{RevModPhys.87.1}. 

To simplify the problem, we transform to the orbital basis that reflects the symmetry of the three-site model. For site $3$, we introduce the spherical harmonics in the usual Condon–Shortley phase convention, $d_m=(d_{x^2-y^2}\pm id_{xy})/\sqrt{2}$. Here, $m=\pm2$ denotes the projection of the atomic orbital moment onto the film normal. Hereinafter, we label these orbitals by the pseudospin $\sigma=m/2$. For sites $1$ and $2$, we utilize the same orbitals rotated by the angle $2\pi/3$ clockwise and counterclockwise, respectively [Fig.~\ref{fig:three_sites}(b)]. In this basis, opposite-orbital hopping is described by a single real matrix element $t_{+-}$ for all pairs of sites. Counterclockwise same-orbital hopping on the orbital $d_{+2}$ is characterized by the amplitude $e^{-2\pi i/3}t_{++}$, where $t_{++}$ is real. Clockwise hopping on this orbital is characterized by the hopping amplitude $e^{2\pi i/3}t_{++}$. The latter is a complex conjugate of the counterclockwise hopping amplitude, as expected for the time-reversed process in a Hermitian system. Since the orbital $d_{-2}$ is a complex conjugate of the orbital $d_{+2}$, the corresponding hopping amplitudes are also related by complex conjugation. Using the Koster-Slater parameters and accounting for the fact that the sign of the hole hopping amplitudes is opposite to that for electrons, we obtain $t_{+-}=0.29$~eV, $t_{++}=0.03$~eV for Ni, and $t_{+-}=0.36$~eV, $t_{++}=0.03$~eV for Co, with a precision of about $0.01$~eV.

\textit{Single-particle states.} The single-particle approximation is a useful starting point for the analysis of the many-particle state. According to group theory~\cite{Dresselhaus2007-br}, the stationary states transform under one of the three cyclic representations of the three-fold rotational symmetry of the three-site model, i.e. they are multiplied by the factor $e^{-2i\pi l/3}$ upon site index cycling $1\to2\to3\to1$, with $l=\pm1,0$. The single-particle g.s. is invariant under index cycling ($l=0$),
\begin{equation}\label{eq:1p_3s_0}
		\psi_{0,s}=\frac{1}{\sqrt{6}}\sum_{n,\sigma}\sigma\hat{c}^{\dagger}_{n,\sigma,s}|0\rangle
\end{equation}
with energy $E_0=-2t_{+-}+t_{++}$.

The lowest-energy chiral single-particle states are
\begin{equation}\label{eq:1p_3s_+}
\begin{split}
	\psi_{+,s}=\frac{\sin\theta_1}{\sqrt{3}}\sum_{n} e^{2\pi in/3}\hat{c}^{\dagger}_{n,+,s}\\
	+\frac{\cos\theta_1}{\sqrt{3}}\sum_{n} e^{2\pi in/3}\hat{c}^{\dagger}_{n,-,s}|0\rangle,
\end{split}
\end{equation}
and 
\begin{equation}\label{eq:1p_3s_-}
	\begin{split}
		\psi_{-,s}=\frac{\cos\theta_1}{\sqrt{3}}\sum_{n} e^{-2\pi in/3}\hat{c}^{\dagger}_{n,+,s}\\
		+\frac{\sin\theta_1}{\sqrt{3}}\sum_{n} e^{-2\pi in/3}\hat{c}^{\dagger}_{n,-,s}|0\rangle
	\end{split}
\end{equation}
characterized by $l=1$ and $-1$, respectively, Here, $\theta_1=\tan^{-1}(1+2t_{++}/t_{+-})\approx 50^\circ$. These states can be interpreted as Bloch waves propagating clockwise and counterclockwise, respectively, around a closed three-site chain. They are related to each other by time reversal, and have the same energy $E_\pm=-t_{+-}-t_{++}$, where we neglect a correction of order $t^2_{++}/t_{+-}$. All the single-particle states are also spin-degenerate.

\textit{Many-particle state.} We expect that many-particle states characterized by populations of $0$, $1$, and $2$ holes per site are relevant to late transition metals due to quantum charge fluctuations. For three sites, the corresponding total populations vary from $0$ to $6$, with very small amplitudes for large deviations from the average populations dictated by the atomic structure.

For Ni, the average d-shell population is somewhat smaller than one hole per site. Thus, the amplitudes of states with one and two holes on three sites are large. However, we do not expect such states to significantly contribute to correlations because hole hopping to the neighboring empty site does not affect interaction energy. Correlations relevant to Ni are expected to stem from the many-hole wavefunction component with one hole per site. This component also has a substantial amplitude in Co populated with somewhat more than one d-hole per site, and is considered throughout the rest of this work. 

Larger hole populations are also expected to contribute to the correlations identified in our work, as outlined in the \textit{Summary}. Quantum charge fluctuations reduce these correlations but are not expected to completely suppress them, as suggested by the persistence of Mott correlations in the metallic state of doped Mott insulators~\cite{doi:10.1126/science.235.4793.1196,PhysRevB.35.8865,https://doi.org/10.48550/arxiv.2206.01063}.
In this work, we focus on identifying the essential form of correlations associated with magnetism, and leave detailed analysis of these effects for future studies.

We consider for simplicity only the FM state with spin polarization normal to the film plane, so the spins of all three holes are either up or down. In the limit of negligible interaction, this state is reduced to the product of three lowest-energy single-particle states Eqs.~(\ref{eq:1p_3s_0})-(\ref{eq:1p_3s_-}) with the same spin. To the lowest order in interaction, its energy is $E_3=-4t_{+-}-t_{++}+2U''/3$, where we use the approximation $\sin\theta_1=\cos\theta_1=1/\sqrt{2}$ for the angle that parameterizes the single-particle states.

We now analyze the interaction-dominated regime using the virtual hopping approximation. The product of three lowest-energy single-particle states Eqs.~(\ref{eq:1p_3s_0})-(\ref{eq:1p_3s_-}) is symmetric with respect to site index cycling, and antisymmetric with respect to orbital reversal. These symmetries are expected to be independent of the interaction magnitude. The wavefunction of the state with these symmetries on the subspace of single-occupancy states is
  \begin{equation}\label{eq:3p_3f}
\begin{split}	
	\psi_{3,s}=\frac{\cos\theta_3}{\sqrt{2}}\sum_{\sigma}\sigma\hat{c}^{\dagger}_{1,\sigma,s}\hat{c}^{\dagger}_{2,\sigma,s}\hat{c}^{\dagger}_{3,\sigma,s}\\
	+\frac{\sin\theta_3}{\sqrt{6}}\sum_{n,\sigma}\sigma\hat{c}^{\dagger}_{n-2,\sigma,s}\hat{c}^{\dagger}_{n-1,\sigma,s}\hat{c}^{\dagger}_{n,-\sigma,s}|0\rangle,
\end{split}
 \end{equation}
where we use cyclic notations for the site index, i.e. $n-1\equiv3$ for $n=1$, and trigonometric parameterization of amplitudes to simplify normalization. Neglecting same-orbital hopping, degenerate second-order perturbation theory gives $\theta_3=30^\circ$ and $E_3=-8t^2_{+-}/U''$.

\begin{figure}
	\includegraphics[width=0.7\linewidth]{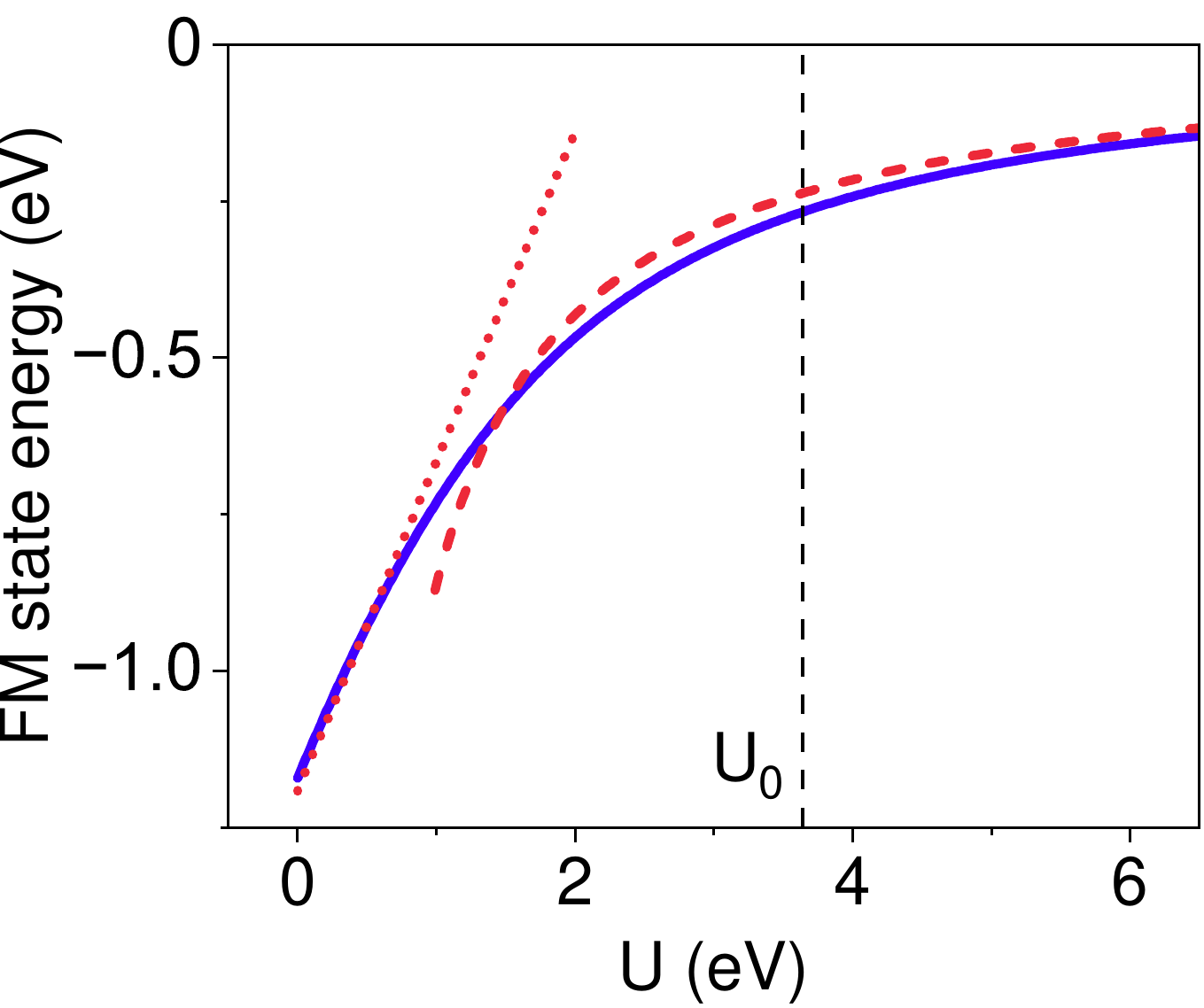}
	\caption{Energy of the FM-ordered state vs $U$ for three holes on three sites, with $J_0=0.21U_0$. Dots: independent-particle approximation, dashed curves: virtual hopping approximation, solid curve: results of exact numeric diagonalization. The hopping parameters of Ni are used in the calculations.}\label{fig:EvsU_3s}
\end{figure}

These analytical approximations are in good agreement with exact numeric diagonalization in the respective limiting regimes, as illustrated in Fig.~\ref{fig:EvsU_3s}. We note that at $U=U_0$, the virtual hopping approximation provides a much better agreement with the exact energy than the small-$U$ approximation, placing the considered system in the strongly correlated regime.

\textit{Two-particle correlations.} We now interpret the three-particle FM state in terms of two-particle correlations. The wavefunction Eq.~(\ref{eq:3p_3f}) with $\theta_3=30^\circ$ can be written as
\begin{equation}\label{eq:3p_3fb}
	\psi_{3,s}=\frac{1}{2\sqrt{6}}\sum_{n,\sigma,\sigma'}\sigma\hat{c}^{\dagger}_{n-2,\sigma,s}\hat{c}^{\dagger}_{n-1,\sigma,s}\hat{c}^{\dagger}_{n,\sigma',s}|0\rangle.
\end{equation}
Consider the component of this state with $n=3$,
\begin{equation}\label{eq:3s_2p_corr}
	\begin{split}
		\psi'_{3,s}=\frac{1}{2}(\hat{c}^{\dagger}_{1,+,s}\hat{c}^{\dagger}_{2,+,s}-\hat{c}^{\dagger}_{1,-,s}\hat{c}^{\dagger}_{2,-,s})\\
		\times(\hat{c}^{\dagger}_{3,+,s}+\hat{c}^{\dagger}_{3,-,s})|0\rangle
	\end{split}
\end{equation}
normalized here to unity so that it can be viewed as a separate wavefunction. This state is a product of the state $d_{xy}$ on site $3$ and an orbitally FM-correlated singlet state on sites $1$ and $2$,
\begin{equation}\label{eq:3s_corr_12}
		\psi_{12,s}=\frac{1}{\sqrt{2}}(\hat{c}^{\dagger}_{1,+,s}\hat{c}^{\dagger}_{2,+,s}-\hat{c}^{\dagger}_{1,-,s}\hat{c}^{\dagger}_{2,-,s})|0\rangle
\end{equation}
consistent with the two-particle correlations illustrated in Fig.~\ref{fig:MKK}(c). Similarly, the components $\psi'_{1,s}$ and $\psi'_{3,s}$ that correspond to $n=1$ and $n=2$ in Eq.~(\ref{eq:3p_3fb}) describe pairwise FM orbital correlations between sites $2$ and $3$, and sites $1$ and $3$, respectively.

To relate these results to the two-site model of Section~\ref{sec:2s}, we first note that the operators in Eq.~(\ref{eq:3s_corr_12}) are defined in the basis of orbitals rotated by $\pm60^\circ$ relative to the standard Condon–Shortley convention. The rotation directions are opposite for sites $1$ and $2$, so these rotations do not affect the operator products in Eq.~(\ref{eq:3s_corr_12}). Using the relation $d_{\pm2}=(d_{x^2-y^2}\pm id_{xy})/\sqrt{2}$, Eq.~(\ref{eq:3s_corr_12}) is then reduced to the first two terms in Eq.~(\ref{eq:spin_2s}). Thus, two-particle intersite orbital correlations in the three-site model are the same as in the two-site model. They can be described as AFM-like in cubic harmonics, and FM-like in the spherical harmonic basis. The latter is more physical since it directly relates to the orbital angular momentum and the effects of SOC discussed in Section~\ref{sec:SOC}.

We note that orbital correlations are distinct from orbital ordering. The latter is associated with unquenched orbital moments. However, orbital moments vanish in the state Eq.~(\ref{eq:3p_3fb}) despite non-vanishing orbital correlations. Orbital quenching in late transition metal ferromagnets is commonly attributed to the lifting of the d-level degeneracy by the effective crystal field~\cite{handley2000modern}. Our results show that orbital quenching can also occur in multi-orbital systems with orbitally unquenched single-particle states, due to the stabilization of the orbital singlet state by the orbital selectivity of hopping. 

\section{Extended system}\label{sec:extended}

In this section, we extend the few-site models of correlations developed in the last two sections to an extended system. We note that the hexagonal symmetry of the triangular lattice does not quench the moments of the single-particle $d_{\pm 2}$ orbitals that dominate the valence states [see Fig.~\ref{fig:d-levels}], so orbital correlations are allowed by the crystal symmetry.

\textit{Virtual hopping approximation.} For realistic hopping and interaction parameters, both two- and three-site model systems discussed above are well-described by the virtual hopping approximation. For an extended system, this implies either an insulating or a "Hund's metal" state characterized by anomalous transport inconsistent with the properties of transition metals~\cite{PhysRevB.102.161118}. However, as discussed in Section~\ref{sec:model}, electronic transport in late transition metal ferromagnets is dominated by the delocalized 4s electrons, without significant anomalous contributions from the quasi-localized 3d electrons.

The appropriateness of the quasi-localized approximation for 3d electrons is implied by the Heisenberg exchange model commonly used to describe magnetism in late transition metals~\cite{handley2000modern}. The AFM Heisenberg Hamiltonian has been rigorously derived for single-band Mott insulators as the second-order virtual hopping correction to the energy of quasi-localized particles~\cite{PhysRevB.2.4302}. In the context of ferromagnetism, this possibility has been demonstrated only in the Kugel-Khomskii model~\cite{Kugel1982}, which is based on the virtual hopping approximation for insulating ferromagnets. Based on these observations, we conclude that virtual hopping likely provides an adequate lowest-order (i.e. second-order in hopping) approximation for the correlations of 3d electrons in extended films.

\textit{Orbital frustration.} As was shown in the previous section, the FM-ordered state of three sites is a superposition of products of an orbital singlet of two sites and a quenched orbital state of the third site. The mechanism of quenching is similar to the frustration of three AFM-coupled Ising spins~\cite{Balents2010-wh}. For the latter system,
only two of the three spins can be AFM-correlated, while the third one becomes frustrated. For the three-site model considered in this work, virtual opposite-orbital hopping of a hole onto the neighboring site occupied by another hole with the same spin is possible if the orbitals of the two holes are FM-correlated in the basis of spherical harmonics whose phase is defined with respect to the direction between two sites. However, the phases of the $d_{\pm2}$ orbitals are rotated by $\pm 120^\circ$ in the direction of the third site, turning constructive opposite-orbital interference (hybridization) into destructive and frustrating FM orbital correlations among three sites. 

For AFM-coupled spins, frustration on the triangular lattice results in the spin liquid state characterized by singlet nearest-neighbor correlations~\cite{Balents2010-wh,RevModPhys.87.1}. Similarly, orbital frustration identified in our model is expected to result in the orbital liquid state of an extended system, which is characterized by the nearest-neighbor two-particle singlet orbital correlations described by Eq.~(\ref{eq:3s_corr_12}), but does not exhibit orbital ordering~\cite{Khaliullin2000-xe}. 

{\it Basis orbitals and Hamiltonian.} Virtual hopping involves hopping of a hole from site $\vec{n}$ onto the neighboring site $\vec{n}+\vec{l}$, and hopping back of the same hole or the hole initially located on that site. For the states projected on the subspace with a single hole per site, this process involves only two neighboring sites, allowing one to introduce site-specific basis orbitals to simplify the analysis of virtual hopping~\cite{RevModPhys.87.1}.

We align the x-axis of the local coordinate system with the direction $\vec{l}$ between two neighboring sites. In the spherical harmonic basis, opposite-orbital hopping amplitude is $t_{+-}$, while same-orbital hopping amplitude is $t_{++}$. We use the virtual hopping approximation -- the second-order degenerate perturbation theory on the subspace of states with a single hole per site -- to determine the effective second-order hopping Hamiltonian

\begin{widetext}
\begin{equation}\label{eq:Heisenberg}
\begin{split}
		\hat{H}^{(2)}=&-\frac{J_1}{2}\sum_{\vec{n},\vec{l},\sigma,\sigma',s,s'}\sigma\sigma'\hat{c}^{\dagger}_{\vec{n},\sigma,s}\hat{c}^{\dagger}_{\vec{n}+\vec{l},\sigma,s}\hat{c}_{\vec{n}+\vec{l},\sigma',s}\hat{c}_{\vec{n},\sigma',s}-\frac{J_2}{2}\sum_{\vec{n},\vec{l},s,|\sigma_1+\sigma_2+\sigma_3+\sigma_4|=1}\hat{c}^{\dagger}_{\vec{n},\sigma_4,s}\hat{c}^{\dagger}_{\vec{n}+\vec{l},\sigma_3,-s}\hat{c}_{\vec{n}+\vec{l},\sigma_2,-s}\hat{c}_{\vec{n},\sigma_1,s}\\
		&-\frac{J_3}{2} \left[\sum_{\vec{n},\vec{l},\sigma,\sigma',s,s'}4ss'\hat{c}^{\dagger}_{\vec{n},\sigma,s}\hat{c}^{\dagger}_{\vec{n}+\vec{l},\sigma,-s}\hat{c}_{\vec{n}+\vec{l},\sigma,-s'}\hat{c}_{\vec{n},\sigma,s'}-\sum_{\vec{n},\vec{l},\sigma,\sigma',s}\hat{c}^{\dagger}_{\vec{n},\sigma,s}\hat{c}^{\dagger}_{\vec{n}+\vec{l},\sigma',-s}\hat{c}_{\vec{n}+\vec{l},-\sigma',s}\hat{c}_{\vec{n},-\sigma,-s}\right],
	\end{split}
\end{equation}
\end{widetext}
where we have neglected corrections of the order $t^2_{++}/t^2_{+-}<0.01$. The summation is over all $\vec{n}$ and $\vec{l}$ so each second-order hopping process is counted twice, $J_1=2t_{+-}^2/U''$, $J_2=t_{+-}t_{++}(2U^2-J^2)/(U^3-UJ^2)$, $J_3=2t_{+-}^2U/(U^2-J^2)$. The operators $\hat{c}_{\vec{n},\sigma}$ annihilate a hole on the corresponding site in the orbital state $\sigma$ defined in the local two-site basis. 

To gain insight into the correlations described by this Hamiltonian, we first consider the subspace of spin-ordered states with spin normal to the film plane. The first term in Eq.~(\ref{eq:Heisenberg}) is the only non-vanishing term on this subspace. The components of this term with $\sigma'=\sigma$ describe Ising FM orbital coupling. The components with $\sigma'=-\sigma$ flip two FM-coupled orbitals of the neighboring sites, which do not have a spin analog because of spin conservation. For orbitals, this is possible because hopping does not conserve orbital angular momentum with respect to a given site. 

The two-orbital flip term prevents orbital ferromagnetism. The g.s. of the Hamiltonian Eq.~(\ref{eq:Heisenberg}) for two sites is an FM-correlated orbital singlet described by Eq.~(\ref{eq:3s_corr_12}) and illustrated in Fig.~\ref{fig:MKK}(c). Since orbital moments on both sites are quenched, such correlations do not prevent pairwise singlet correlations with other sites, allowing the description of the entire extended system in terms of two-site correlations. This state can be interpreted as an orbital version of resonating valence bond (RVB) state, or equivalently as an orbital liquid~\cite{Khaliullin2000-xe}. 

{\it Energy of orbital correlations.} To gain further insight into the correlations described by Eq.~(\ref{eq:Heisenberg}), we first assume a state with completely disordered site spins and orbital moments. A pair of neighboring sites $\vec{n}$, $\vec{n}+\vec{l}$ can then be described by a $16\times16$ density matrix which accounts for two possible orbital and spin directions on each site, and in the completely disordered state is simply proportional to the unity matrix,
\begin{equation}\label{eq:rho_uncorr}
\hat{\rho}_{\vec{n},\vec{n}+\vec{l}}=\frac{1}{16}\sum_{\sigma,\sigma',s,s'}\hat{n}_{\vec{n},\sigma,s}\hat{n}_{\vec{n}+\vec{l},\sigma',s'}.
\end{equation}
The contribution to energy due to virtual hopping between these two sites is
\begin{equation}\label{eq:E_uncorr}
	E_{\vec{n},\vec{n}+\vec{l}}=Tr(\hat{\rho}_{\vec{n},\vec{n}+\vec{l}}\hat{H}^{(2)})=-t^2_{+-}(\frac{1}{U}+\frac{1}{2U''}).
\end{equation}
Accounting for the fact that each site has six nearest neighbors, the total energy of the uncorrelated state of an $N$-site system is $3NE_{\vec{n},\vec{n}+\vec{l}}$.

The FM spin-ordered, orbitally uncorrelated state can be similarly described by the $4\times 4$ density matrix accounting for two possible orbital states of each site,
\begin{equation}\label{eq:rho_s}
	\hat{\rho}_{\vec{n},\vec{n}+\vec{l}}=\frac{1}{4}\sum_{\sigma,\sigma',s}\hat{n}_{\vec{n},\sigma,s}\hat{n}_{\vec{n}+\vec{l},\sigma',s}.
\end{equation}
The energy of this state for the $N$-site system determined as in Eq.~(\ref{eq:E_uncorr}) is $-3Nt_{+-}^2/U''$. It is easy to see that this energy is higher than that of the completely disordered state. Spin ordering without orbital correlations raises energy instead of lowering it because some of the virtual hopping channels are eliminated due to the Pauli exclusion principle.

The FM spin-ordered, orbitally correlated state of two neighboring sites described by the wavefunction Eq.~(\ref{eq:3s_corr_12}) is an eigenstate of $\hat{H}^{(2)}$ with energy $-8t_{+-}^2/U''$. The corresponding energy of the extended system is $-24Nt_{+-}^2/U''$. It is lower than that of the uncorrelated state by $\Delta E_{corr}=3Nt^2_{+-}(15U-2U'')/2UU''$. Thus, orbital correlations are necessary for spin ordering in the considered virtual-hopping approximation. This approximation underestimates the effects of quantum fluctuations and deviations from the average population of one d-hole per site, and neglects the effects of hybridization with 4s-electrons. Because of these limitations, our model cannot quantitatively account for the energy of the magnetic system. Nevertheless, it elucidates the nature of correlations and their role in magnetic ordering.

\section{Effects of SOC}\label{sec:SOC}

Orbital moments are quenched in the orbital liquid state, making it more challenging to observe this state than orbital ordering. Nevertheless, orbital correlations may be manifested in phenomena that involve SOC, such as magnetic anisotropy, anomalous and spin Hall effects~\cite{RevModPhys.82.1539}. In this section, we show that orbital correlations identified above should result in a strong dependence of magnetic anisotropy~\cite{PhysRevB.39.865} on the relation between Mott-Hund's interaction and hopping amplitudes, providing a simple verifiable prediction for our model.

Thin films of late transition metal ferromagnets commonly exhibit an anomalously large PMA~\cite{Johnson1996-jk}, which is particularly important for modern magnetic nanodevices~\cite{Poletkin2022-hd} and is usually interpreted in terms of single-particle orbital hybridization across the interface~\cite{Bruno1989-zn}. Atomistically, magnetic anisotropy results from SOC-mediated spin coupling to orbital moments. At the single-particle level, the latter are controlled by the effective crystal field effects, as illustrated in Fig.~\ref{fig:d-levels} for the system considered in this work. On the other hand, observation of two-ion anisotropy suggests that in some cases, a single-particle approximation may be inadequate~\cite{PhysRevB.83.024404}. We now analyze such many-particle effects using the Hubbard Hamiltonian Eq.~(\ref{eq:Hunds}) including the SOC term. 

The SOC Hamiltonian has a simple form in the basis of spherical harmonics $d_{\pm2}$ because the latter are not mixed by the orbital moment operators,
\begin{equation}\label{eq:SOC}
\hat{H}_{SOC}=2\lambda \sum_{\vec{n},\sigma,s} s\cdot\sigma \hat{n}_{\vec{n},\sigma,s},
\end{equation}
with $\lambda\approx 40$~meV for Ni~\cite{PhysRevLett.101.236404}. 

We use the two-site model as an approximation for the two-site correlations in the extended system, as discussed in the previous section. Including the SOC term in the Hamiltonian, we obtain the two-site FM g.s. wavefunction for spin $s$ normal to the film,
\begin{equation}\label{eq:psi_2s_lambda}
	\begin{split}
		\psi_{t,s}&=\cos\theta_t(\sin\phi\hat{c}^{\dagger}_{1,-,s}\hat{c}^{\dagger}_{2,-,s}-\cos\phi\hat{c}^{\dagger}_{1,+,s}\hat{c}^{\dagger}_{2,+,s})\\
		&-\frac{\sin\theta_t}{\sqrt{2}}(\hat{c}^{\dagger}_{1,-,s}\hat{c}^{\dagger}_{1,+,s}-\hat{c}^{\dagger}_{2,-,s}\hat{c}^{\dagger}_{2,+,s})|0\rangle.
	\end{split}
\end{equation}
At $\phi=\pi/4$, this wavefunction reduces to the spin-triplet state derived in Section~\ref{sec:2s} in the limit of negligible SOC, with energy $E_t=\frac{U''}{2}-\sqrt{(\frac{U''}{2})^2+4t^2_{+-}}$ and $\theta_t=\tan^{-1}(-E_t/2t_{+-})$.

The deviations of $\phi$ from $\pi/4$ due to SOC break the symmetry between the term with two up orbitals and the term with two down orbitals, the only terms in Eq.~(\ref{eq:psi_2s_lambda}) that contribute to SOC energy. To the lowest order in $\lambda/t_{+-}$, $\phi=\pi/4-4s\lambda/E_t$,
and the energy is $E=E_t+E_{SOC}$, where
\begin{equation}\label{eq:E_SOC}
	E_{SOC}=\frac{8\lambda^2}{E_t}(1+\frac{E_t}{U''-2E_t})
\end{equation}
is the contribution of SOC to the two-particle state energy. Analysis omitted here for brevity shows that the effect of SOC on the in-plane spin-polarized state is two orders of magnitude smaller because the subspace of orbitals $d_{\pm2}$ includes only orbital moments normal to the film. This small energy is associated with the higher-order SOC-induced effects that involve two-hole spin-orbit correlations. Neglecting this correction, the energy of the normal-spin triplet components is smaller than that of the in-plane spin-polarized state by $-E_{SOC}$, which describes the two-site PMA energy. 

\begin{figure}
\includegraphics[width=1.0\linewidth]{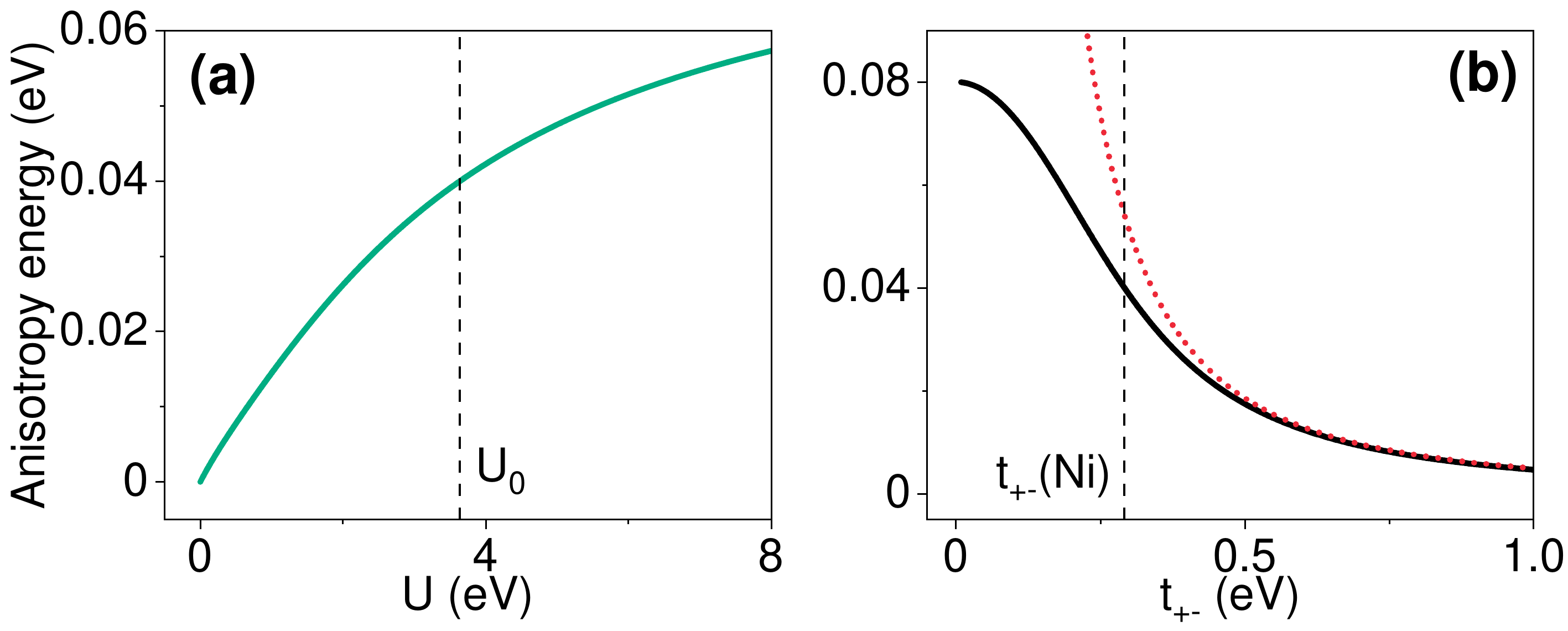}
\caption{(a) Two-site PMA energy of the two-site model vs $U$ calculated using the hopping parameters of Ni, $\lambda=40$~meV, and $J_0=0.21U_0$. (b) Same is (a) vs $t_{+-}$, using $U''=2.87$~eV. Solid curve: numerical solution, dashed curve: analytical approximation represented by Eq. \ref{eq:E_SOC}.}\label{fig:PMA}
\end{figure}

The effect of correlations on magnetic anisotropy is illustrated in Fig.~\ref{fig:PMA}(a), which shows the dependence of the two-site anisotropy energy on the onsite interaction for the hopping parameters of Ni and $\lambda=40$~meV. The anisotropy energy increases with increasing $U$, since a large interaction confines the two holes to the respective sites, increasing the amplitudes of wavefunction components that contribute to SOC.

Since the onsite Mott-Hund's interaction is mainly a property of the atomic species, a more experimentally relevant prediction of our model is the increase of magnetic anisotropy with decreasing hopping parameter $t_{+-}$ at fixed interaction [Fig.~\ref{fig:PMA}(b)], which is explained by the same mechanism as the dependence on $U$. Experimentally, hopping can be varied by introducing impurities with electronic level structure mismatched with that of late transition metal host, such as B, Si or Al~\cite{Bozec2000-nc}, which may allow one to judiciously control PMA in nanodevices based on ultrathin magnetic films.

\section{Summary}\label{sec:discussion}

In this work, we used the two-orbital Hubbard-Kanamori model including spin-orbit coupling to analyze correlations in ultrathin films of late transition metal ferromagnets such as Ni and Co. Our analysis suggests that the wavefunction of d-electrons hosting magnetism is not well-approximated by the Slater determinant of single-particle states. Instead, it can be well-described by the virtual hopping approximation, tentatively placing these systems in the strongly-correlated regime in the sense that interactions dominate the many-particle wavefunction of d-electrons.

Our main result is the prediction that correlations take the form of the orbital liquid state characterized by ferromagnetic orbital correlations between neighboring sites, but vanishing onsite orbital moments in the limit of negligible SOC. We show that in contrast to Kugel-Khomsii materials, orbital ordering is prevented in the considered systems by the mismatch between orbital and crystal symmetries, resulting in orbital frustration. Because of orbital moment quenching, the orbital liquid state is significantly more challenging to observe than the orbitally-ordered state. Nevertheless, it is expected to be manifested in phenomena involving spin-orbit coupling such as magnetic anisotropy, spin Hall and anomalous Hall effects~\cite{RevModPhys.82.1539}, and Dzyaloshinskii-Moriya interaction~\cite{PhysRev.120.91}. In particular, we showed that the correlations may contribute to the commonly observed enhancement of perpendicular magnetic anisotropy in ultrathin films. The significance of this contribution can be experimentally tested by suppressing hopping via doping with impurities such as B, Si, or Al.

Our model focuses on the component of the many-electron wavefunction characterized by the d-level population of one hole per site, whose amplitude is maximized in the Co$_{50}$Ni$_{50}$ alloy. However, orbital correlations identified in our work are expected to persist in thin films with larger d-hole populations. Consider, for instance, a thin-film fcc Co$_{50}$Fe$_{50}$ alloy that contains about two 3d holes per site. According to the Hund's rules, their spins are expected to be aligned, in agreement with the Slater-Pauling curve~\cite{handley2000modern}. The second Hund's rule can be satisfied if the holes occupy, for example, orbitals $d_{+2}$ and $d_{+1}$ to maximize their total orbital moment. Their opposite-orbital hopping onto the nearest neighbor site is then maximized if the holes on that site occupy states $d_{-2}$ and $d_{-1}$.

Following the same arguments as for a single hole per site, opposite-orbital hopping stabilizes the orbital singlet state of two sites, while orbital frustration on the fcc lattice stabilizes an orbital liquid state. The same argument extends to states with three holes on two sites, with one hole shared between the sites. It starts to break down for populations exceeding two holes per site, since the same state $d_{0}$ must be occupied on both sites, whose energy can be lowered by AFM spin correlations of the two holes, instead of FM correlation. This is consistent with the observation that the largest moment per site achievable in transition metal alloys is about $2.5\mu_B$~\cite{handley2000modern}, and with antiferromagnetism of Fe monolayers and FeMn alloys~\cite{PhysRevLett.94.087204,handley2000modern}.

In bulk Co and Ni, each site has six nearest neighbors in the two (111) planes above and below the plane containing this site and its six neighbors considered in our work. In the spherical harmonic basis, same-orbital hopping is dominant for the out-of-plane neighbors, resulting in AFM two-site orbital correlations. The picture of orbital correlations then becomes significantly more complicated than that in ultrathin films. Nevertheless, since the orbital singlet is characterized by a vanishing onsite orbital moment, FM in-plane orbital correlations are compatible with AFM out-of-plane correlations, allowing for an orbital liquid state in the bulk materials. We leave its detailed analysis for future studies.

Our approach can be extended to other magnetic systems, such as small clusters of metals~\cite{clusters1,clusters2}, layered oxides such as $\text{LiVO}_2$ and $\text{NaTiO}_2$~\cite{KK1,KK2}, and two-dimensional ferromagnets such as vanadium and chromium trihalides~\cite{CrI1,CrI2}. For the latter, orbital degrees of freedom are believed to be important, but their specific role remains elusive. In particular, the magnetic moment of almost exactly $3\mu_B$ per Cr atom is consistent with the spin alignment of its three d-electrons, without any substantial contribution of orbital moment. The Cr atoms form a 2d hexagonal lattice whose symmetry prevents orbital ordering, as discussed above for the triangular lattice. We conclude that d-electrons in chromium trihalides likely form an orbital liquid characterized by singlet orbital correlations dictated by the Hund's rules that stabilize long-range spin ordering and result in negligible orbital moment.

\section*{Acknowledgments}
Two-site analysis by SU was supported by the DOE BES Award \# DE-SC0018976. The rest of the work was supported by the NSF Awards ECCS-2005786 and ECCS-1804198.

\bibliography{F}
\bibliographystyle{apsrev4-2}
\end{document}